\documentclass[12pt,preprint]{aastex}

\usepackage{amssymb}
\usepackage{amsmath}
\usepackage{graphicx}
\usepackage{psfig}

\def\lesssim{\mathrel{\hbox{\rlap{\hbox{\lower4pt\hbox{$\sim$}}}\hbox{$<$}}}}
\def\gtrsim{\mathrel{\hbox{\rlap{\hbox{\lower4pt\hbox{$\sim$}}}\hbox{$>$}}}}

\newcommand{\mbfB}{\mathbf{B}}
\newcommand{\mbfE}{\mathbf{E}}

\newcommand{\mbfv}{\mathbf{v}}

\newcommand{\mbfz}{\mathbf{z}}

\newcommand{\mbfj}{\mathbf{j}}
\newcommand{\mbfk}{\mathbf{k}}

\newcommand{\mbfnabla}{\mathbf{\nabla}}
\newcommand{\mbfphi}{\mathbf{\phi}}

\newcommand{\omc}{\omega_c}

\newcommand{\omci}{\omega_{ci}}
\newcommand{\phidot}{\dot\phi_1}
\begin{document}

\title{The Weak Field Limit of the Magnetorotational Instability}

\author{Julian H. Krolik\altaffilmark{1}}
\author{Ellen G. Zweibel\altaffilmark{2}}
\altaffiltext{1}{Johns Hopkins University, Baltimore, MD U.S.A.}
\altaffiltext{2}{U Wisconsin-Madison, 475 N Charter St, Madison,
                 WI 53706, U.S.A.}

\shortauthors{Zweibel \& Krolik}
\shorttitle{Weak Field Instability}

\begin{abstract}
We investigate the behavior of the magneto-rotational instability in the limit
of extremely weak magnetic field, i.e., as the ratio of ion cyclotron frequency
to orbital frequency ($X$) becomes small.  Considered only in terms of cold
two-fluid theory, instability persists to arbitrarily small values of $X$,
and the maximum growth rate is of order the orbital frequency except
for the range $m_e/m_i < |X| < 1$, where it can be rather smaller.
In this range, field aligned with rotation ($X > 0$) produces slower
growth than anti-aligned field ($X < 0$).  The maximum growth
rate is generally achieved at smaller and smaller wavelengths as $|X|$
diminishes.  When $|X| < m_e/m_i$, new unstable 
``electromagnetic-rotational"
modes appear that do not
depend on the equilibrium magnetic field.  Because the most rapidly-growing
modes have extremely short wavelengths when $|X|$ is small, they are often
subject to viscous or resistive damping, which can result in suppressing
all but the longest wavelengths, for which growth is much slower.
We find that this sort of damping is likely to curtail severely the
frequently-invoked mechanism for cosmological magnetic field growth in
which a magnetic field seeded by the Biermann battery is then amplified
by the magneto-rotational instability.  On the other hand, the small
$|X|$ case may introduce interesting effects in weakly-ionized disks
in which dust grains carry most of the electric charge.

\end{abstract}

\section{Introduction}\label{s:introduction}

    Since its reintroduction into astrophysics by Balbus \& Hawley (1991),
the magneto-rotational instability (MRI) has become an essential ingredient
in our understanding of astrophysical fluids in a state of differential
rotation.  As a result of its action, strong turbulence is created anywhere
there is a weak magnetic field in a conducting fluid whose rotation
rate decreases away from the rotation axis.  The turbulent magnetic (and,
to a lesser degree, fluid) stress it creates is the most likely source of
angular momentum transport in accretion disks (Balbus \& Hawley 1998).
The joint action of turbulence and differential rotation can drive
a dynamo capable of maintaining the magnetic energy at a few percent
of the fluid internal energy.  Applications of these mechanisms have
been found in a wide range of circumstances, from supernovae (Akiyama
et al. 2003) to galaxies (Kim et al. 2003).

    Somewhat paradoxically, it is a condition for growth that the field
be weak.  To be precise, when gas and magnetic pressure combine to
support the disk material against the vertical component of gravity,
the matter's scale-height is $h \sim \sqrt{v_A^2 + c_s^2}/\Omega$,
where $v_A$ is the Alfven speed, $c_s$ the sound speed, and $\Omega$
the local orbital frequency.  Because magnetic tension suppresses growth
for wavelengths $\lambda \lesssim 2\pi v_A/\Omega$, only if the magnetic
energy density is less than the gas pressure (i.e., the plasma
$\beta \equiv c_s^2/v_A^2 > 1$) can modes with short enough wavelength
to fit in the disk ($\lambda \lesssim h$) be unstable.  At the same time,
although the magnitude of the fastest growing wavelength depends
(linearly) on the field strength, its growth rate 
$\sim \Omega$ is entirely independent
of the field's intensity.  On the other hand, if there were truly zero
field, magnetic dynamics would, of course, be irrelevant.  Thus, it
appears that the nature of the MRI in the limit of progressively weaker
field is ill-defined, at least within the framework of ideal MHD.

    It is the goal of this paper to begin clarifying what happens in
this extreme weak-field limit, developing linear theory to account
both for growth and damping.  We do so in part to illuminate this point of
principle, but we also identify several astrophysical contexts
in which the field may be so weak that the special effects we discover
in the weak-field limit may be of interest.  These
include galaxies so young that their interstellar magnetic fields have
not yet grown to interesting amplitude, accretion systems in primordial
galaxies where the seed fields may have been very weak, and proto-stellar
accretion flows in which the charge resides primarily on massive grains.

Some aspects of our work are related to results already in the literature.
In particular, there has already been significant study of how the MRI
operates in media where the magnetic field is weak in a different sense:
where the field is considered weak not because it fails to dominate gravity,
but because it is not strong enough for the Lorentz force it produces to
dominate the momentum transfer associated with collisions between charged
species and neutrals.  This case is called the Hall regime, and has
been analyzed by Balbus \& Terquem (2001) and Salmeron \& Wardle (2003).
We uncover Hall effects too, but of a different sort (see
\S~\ref{s:threeregimes}).

Similarly, we will be interested in a breakdown of ideal MHD in the sense
that the electrons and ions are not tied perfectly to the magnetic field.
A different failure of the MHD approximation in the context
of the MRI has been examined by Quataert, Dorland, \& Hammett (2002), 
Sharma, Hammett, \& Quataert (2003), and Balbus (2004), who have explored 
the implications of collisionless behavior.  They find that anisotropic 
pressure, anisotropic viscosous transport, and
Landau damping, none of which is captured by MHD, can be
important.
%Although collisionless effects are potentially significant 
%in the short wavelength regime that must be considered for weak magnetic
%fields, both identically vanish for the situation considered in this
%paper---a point we discuss further
In \S~\ref{s:damping}, we argue that all three of these effects are
unimportant in the weak-field regime.  Finally, Kitchatinov
\& R\"udiger (2004) have considered the MRI in protogalactic disks and
estimated the fieldstrength below which the instability is resistively damped.
 
\section{Two-Fluid Formulation and Derivation of the Dispersion Relation}
\label{s:formulation}

To explore this new regime, we adopt the simplest non-trivial model that
can describe the physics of the weak-field regime: a two-fluid picture of
a cold but fully-ionized disk, threaded by a uniform magnetic field
$\hat\mbfz B$,
orbiting in a gravitational potential determined by external sources.
The conventional single-fluid MHD limit implicitly assumes that the
plasma and the fieldlines are frozen together; in order to see how they
decouple in the limit of weak field, we must follow the dynamics of the
positive and negative charge-carriers separately.   That is to say,
we define the system in terms of separate force equations for two
incompressible fluids, with no pressure gradients.  The fields are
determined by using the currents derived from the velocities of the
two charged fluids in the Amp\`ere Equation and using the changing
magnetic fields created by those currents in Faraday's Law.

We use cylindrical polar coordinates $(r,\phi,z)$ and denote radial derivatives
by primes.  In the equilibrium state, particles follow circular orbits of
radius $r_0$ with angular velocity $\dot\phi_0\equiv\Omega$ given by the
usual expression $\Omega(r_0)=[r_0^{-1}V^{'}(r_0)]^{1/2}$.  The particles are
prevented from gyrating around the magnetic fieldlines by a radial electric
field $\mbfE=-r_0\Omega\hat\mbfphi\times B\hat\mbfz/c$.

We consider small axisymmetric perturbations of this system, denoting
the electric and
magnetic field perturbations by $\mbfE_1$ and $\mbfB_1$ and the perturbations
of particle radius and angular velocity by $r_1$ and $\phidot$.  In the cold
plasma approximation, all the particles of a given species have the same
perturbed orbits, but the orbits differ between species.  We assume all
perturbed quantities depend on time and vertical coordinate $z$ as
$\exp{(\sigma t + ikz)}$ and perform a local analysis in which radial
derivatives of the perturbed quantities are dropped.

The linearized equations of motion for particles of charge $q$ and mass $m$
are
\begin{equation}\label{e:rmotion}
\sigma^2 r_1=r_1\Omega^2+2r_0\Omega\phidot -r_1V^{''}+\frac{q}{m}E_{1r}+\omc r_0\phidot,
\end{equation}
\begin{equation}\label{e:tmotion}
r_0\sigma\phidot=-2\sigma r_1\Omega+\frac{q}{m}E_{1\phi}-\omc\sigma r_1,
\end{equation}
where $\omc\equiv qB/mc$ is the cyclotron frequency (note that the sign of
$\omc$ depends on the signs of both $q$ and $B$).  In deriving these equations
we have assumed $B_{1z} \equiv 0$, which is consistent with the local analysis,
and we have used $\mbfE_0+\mbfv_0\times\mbfB/c\equiv 0$.

We solve equations (\ref{e:rmotion}) and (\ref{e:tmotion}) for the perturbed
velocity components $\sigma r_1$ and $r_0\phidot$ in terms of $\mbfE_1$.
The results are
\begin{equation}\label{e:r1}
\sigma r_1=\left(\frac{q}{m}\right)\frac{\sigma E_{1r}+\left(\omc+2\Omega\right)E_{1\phi}}{\omc^2+4\Omega\omc+\kappa^2+\sigma^2},
\end{equation}
\begin{equation}\label{e:t1}
r_0\phidot=\left(\frac{q}{m}\right)\frac{\left(\sigma+\frac{2r_0\Omega\Omega^{'}}{\sigma}\right)E_{1\phi}-\left(\omc+
2\Omega\right)E_{1r}}{\omc^2+4\Omega\omc+\kappa^2+\sigma^2},
\end{equation}
where $\kappa^2\equiv 4\Omega^2+2r\Omega\Omega^{'}$ is the squared epicyclic
frequency. Although our analysis up to this point has been valid for
any gravitational potential, from this point on,
%because are interested primarily in weak magnetic field effects,
we assume a Keplerian rotation law, for which $\kappa^2=\Omega^2$ and
$2r\Omega\Omega^{'}=-3\Omega^2$.

The perturbed current density $\mbfj_1$ is related to the perturbed velocity
components by
\begin{equation}\label{e:j1}
\mbfj_1(\mbfE_1)=\sum_{\alpha}n_{\alpha}q_{\alpha}\mbfv_{\alpha}.
\end{equation}
The summation is over species, and $n$ is number density. We assume two
species, of charge $\pm e$ and equal density. The dependence of $mbfj_1$
on $\mbfE_1$
follows from eqns. (\ref{e:r1}) and (\ref{e:t1}). On the other hand, $\mbfj_1$
is also related to $\mbfE_1$ through Faraday's law and Amp\'ere's law, with
the displacement current dropped
\begin{equation}\label{e:jE}
\mbfj_1(\mbfE_1)=-\frac{c^2k^2}{4\pi\sigma}\mbfE_1.
\end{equation}

Combining eqns. (\ref{e:j1}) and (\ref{e:jE}) and using eqns. (\ref{e:r1}) and
(\ref{e:t1}) gives a dispersion relation, which we write in dimensionless form
using the variables
\begin{equation}\label{e:vars}
\nu\equiv\frac{\sigma}{\Omega},\;\;Z^2\equiv\frac{4\pi ne^2}{c^2k^2m_i},\;\;X\equiv\frac{\omci}{\Omega},\;\;
R\equiv\frac{m_i}{m_e}.
\end{equation}
The dimensionless growth rate $\nu$ is thus measured in units of the
orbital frequency, while the dimensionless wavelength is given by $Z$.
The quantity $4\pi n e^2/m_i$ in the definition of $Z$ is the squared ion
plasma frequency, $\omega_{pi}^2$, so $Z^2 = \omega_{pi}^2/k^2c^2$.  Thus,
$Z$ measures the wavelength in terms of the ion inertial length
$\delta_i\equiv c/\omega_{pi}$: $Z = 1/(k\delta_i)$.
One interpretation of $\delta_i$ is that it is
the Larmor radius of an ion traveling at the Alfv\'en speed: that is,
$v_A=\delta_i\omci$. In an electron-proton plasma, $\delta_i\sim 2.2\times 10^7
n^{-1/2}$ cm.  In most astrophysical situations this
is much less than the size of the system, so global modes have $Z\gg 1$.
Note that with these definitions, $X^2/Z^2=k^2v_A^2/\Omega^2$.

Field strength is described in terms of $|X|$\footnote{Note that our
definition of $X = 2 x^{-1}(\mu_e/m_i)$, where $x$ is the Hall parameter defined in
Balbus \& Terquem 2001 and $\mu_e/m_i$ is the ratio of the mass per electron to
the ion mass.   In a fully-ionized plasma $\mu_e/m_i \simeq 1$, but in a state of
weak ionization this ratio could be much larger.}.  In the standard theory of
the MRI, $\vert X\vert\gg 1$, but we will consider the full range of
$\vert X\vert$.  In most of the paper we will set $R$ equal to its value
in an electron-proton plasma, but in \S~\ref{s:equalmass} we will
also consider plasmas with equal mass charge carriers, for which $R=1$.

The dispersion relation can be written compactly in terms of the functions
\begin{equation}\label{e:F}
F_{\nu}\equiv\frac{1}{X^2+4X+1+\nu^2}+\frac{R}{R^2X^2-4RX+1+\nu^2},
\end{equation}
and
\begin{equation}\label{e:G}
G_{\nu}\equiv\frac{2+X}{X^2+4X+1+\nu^2}+\frac{R(2-RX)}{R^2X^2-4RX+1+\nu^2}.
\end{equation}
In terms of these two functions, it is
\begin{equation}\label{e:DR}
\left[1+Z^2(\nu^2-3)F_{\nu}\right]\left(1+Z^2\nu^2F_{\nu}\right)+Z^4\nu^2G_{\nu}^2=0.
\end{equation}

Much of the subsequent analysis in this paper is based on eqn.~(\ref{e:DR}).
It is an 8th degree polynomial equation for $\nu^2$.  In most of the parameter
space that interests us, we find that the solutions are positive and negative
pairs of either purely real or purely imaginary numbers. When this is so,
solutions of
eqn.~(\ref{e:DR}) with $\nu^2 > 0$ correspond to instability, and perturbations are either
purely oscillatory or purely exponential in time.

It is likewise clear from the form of eqn.~(\ref{e:DR}) that the
magnetic field strength enters only in terms of the parameter $X$.
However, $X$ appears sometimes by itself (representing ions) and sometimes in a product
with $R$ (representing electrons).  Thus, whether the field is strong or weak depends on
whether the cyclotron frequency (usually of the positive charge
carrier) is large or small compared to the orbital frequency.

\section{Analysis of the Dispersion Relation}\label{s:results}

\subsection{Three regimes of $|X|$: $\gg 1$, $\sim 1$, and $\ll 1/R$}
\label{s:threeregimes}

We first consider the conventional limit of eqn.~(\ref{e:DR}): the
case of ordinary MHD, in which $X\gg 1$ and the ratio $R$ is that of
an electron-proton plasma, 1837.  In this limit,
the functions $F_{\nu}$ and $G_{\nu}$ take the asymptotic forms
$F_{\nu}\rightarrow X^{-2}$, $G_{\nu}\rightarrow -2X^{-2}$.
Equation~(\ref{e:DR}) becomes
\begin{equation}\label{e:MHD}
\nu^4 + \nu^2\left(1+2\frac{X^2}{Z^2}\right)+\frac{X^2}{Z^2}\left(\frac{X^2}{Z^2}-3\right)=0.
\end{equation}
Equation (\ref{e:MHD}) is the standard MRI dispersion relation for a Keplerian
disk, albeit written in an unfamiliar form. There is a positive real root when
$X^2/Z^2 = k^2 v_A^2/\Omega^2 < 3$.
As has been well-known since their original paper (Balbus \& Hawley 1991),
the maximum growth rate occurs for $k^2v_A^2/\Omega^2=15/16$, and
$\nu_{max}$ = 0.75.  Thus, the maximum growth rate occurs for wavelengths
$\simeq 1.8$ times longer than the minimum wavelength for growth, and
$\nu \propto k$ for wavelengths much larger than the minimum.

In the MHD regime, both the ions and electrons closely follow the fieldlines,
which have the $\mbfE\times \mbfB$ velocity $c\mbfE\times\mbfB/B^2$ .
As $|X|$ decreases and $\omci$ approaches $\Omega$, the ions deviate from the
fieldlines, while the electrons, due to their much smaller mass, continue
to follow them.  This regime of behavior (i.e., $|X| \sim 1$ or less) is
often called the ``Hall regime" of the MRI because charge-sign differences
lead to noticeable dynamical contrasts.  As we shall see, that is definitely
the case here, but with the slight interpretive shift that what really matters
is not so much charge-sign {\it per se} as the sense of the
Larmor orbit for the heavier particle relative to the sense of the
gravitational orbits for all particles.

Much of the essential behavior of this regime is captured by approximating
$F_{\nu}$ and $G_{\nu}$
%as defined in eqns. (\ref{e:F}) and (\ref{e:G})
by $F_{\nu}\rightarrow (X^2+4X+1+\nu^2)^{-1}$, and
$G_{\nu}\rightarrow -(2X+1+\nu^2)F_{\nu}/X$.  The resulting dispersion
relation, which is valid as long as $|RX| \gg 1$ but $|X|$ is
not $\gg 1$, is
\begin{equation}\label{e:postMHD}
\nu^4 + \nu^2\left(1+2\frac{X^2}{Z^2}+\frac{X^2}{Z^4}\right)+\frac{X^2}{Z^2}
\left(\frac{(X^2+4X+1)}{Z^2}-3\right)=0.
\end{equation}
Equation (\ref{e:postMHD}) predicts instability for $(X^2+4X+1)/3 < Z^2$.
This can occur in either of two ways: In the special range of $X$ given
by $-2-\sqrt{3} < X < -2+\sqrt{3}$, $X^2+4X+1 < 0$, and the 
system is unstable at all wavelengths.  That is, when the Larmor frequency
and the orbital frequency are comparable in magnitude but have opposite
sign for the heavier charged particle, unstable growth occurs at {\it all}
wavelengths.  Alternatively, whenever $X$ is in the range for which the
approximations leading to eqn. (\ref{e:postMHD}) are valid, growth
occurs, but only at sufficiently long wavelengths.  Because
$|X| \sim 1$, the quadratic character of the expression
for the minimum unstable wavelength leads to significant differences
between positive and negative $X$: it is longer for positive $X$ and
shorter for $X<-2-\sqrt{3}$.  Similarly, the maximum growthrate is $\sim O(1)$
for negative $X$, whereas it can be substantially smaller for positive $X$.

When $|X|$ is still smaller, $\ll 1/R$, a different analytic approximation
applies: $F_\nu \rightarrow R/(1+\nu^2)$ and $G_\nu \rightarrow 2R/(1+\nu^2)$.
In that limit, the dispersion relation becomes
\begin{equation}\label{e:vsmallx}
\nu^4 (1 + s)^2 + \nu^2\left(2 - s + s^2\right) + (1 - 3s) = 0,
\end{equation}
where $s \equiv RZ^2$ is proportional to the square of the
wavelength in units of the electron
inertial length $\delta_e\equiv c/\omega_{pe}$: $s=(k^2\delta_e^2)^{-1}$ is
the natural analog of $Z^2$ in a regime where the ions play no role in the
dynamics and the mode is carried by electrons.  Eqn.~(\ref{e:vsmallx}) has
one solution corresponding to a mode that grows unstably when $s > 1/3$:
$\nu = \sqrt{3s-1}/(1+s)$.
Even in this regime, the growth rates can be comparable to the
rotation rate: the peak growth rate, $\nu\sim 0.75$, occurs for $s\sim 1.65$.
However, because $\delta_e\sim 5\times 10^5 n^{-1/2}$~cm is so small in
most systems of interest, we are primarily interested in the case $s\gg 1$.
In this limit, $\nu\sim\sqrt{3/s}\ll 1$.  That is, the modes of greatest
interest grow very slowly.

\subsection{Exact numerical solutions}

\begin{figure}
\centerline{\psfig{file=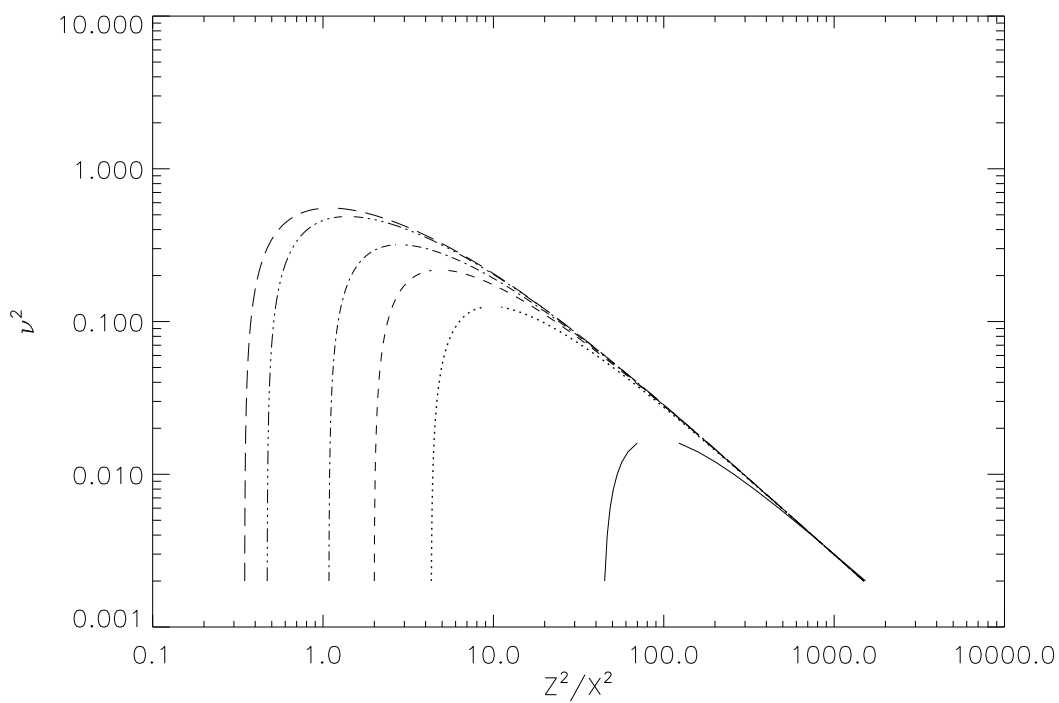,angle=90,width=5.5in}}
\caption{\label{fig:posdisperse}Squared growth-rate as a function of scaled
wavelength when $X>0$.  The linestyle identifications are: $|X|=0.1$: solid;
$|X|=0.5$: dotted; $|X|=1.0$: dashed; $|X|=2.0$: dot-dashed; $|X|=10.0$:
triple-dot-dashed; $|X|=100.$: long-dashed.}
\end{figure}

\begin{figure}
\centerline{
\psfig{file=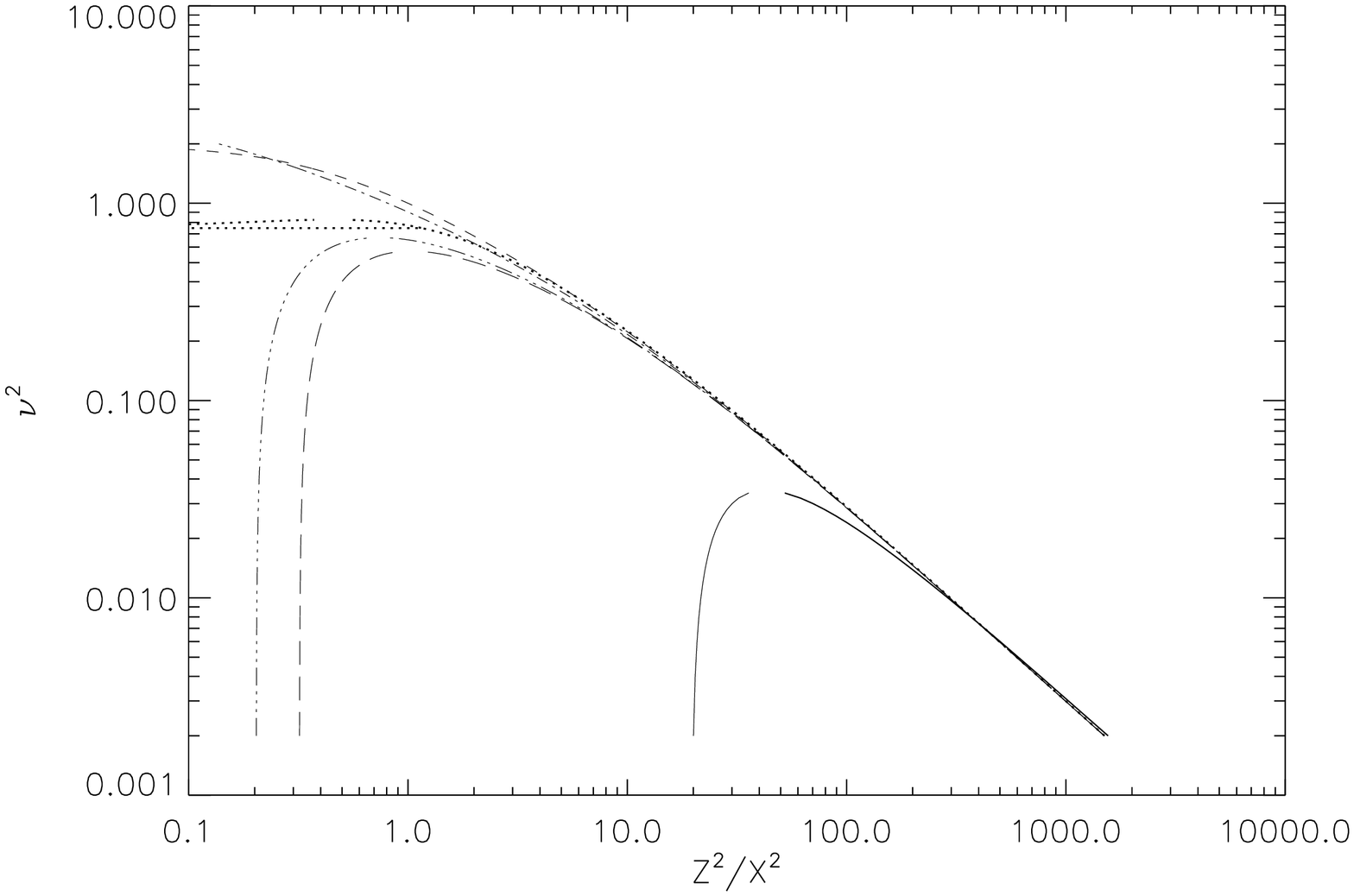,angle=90,width=5.5in}}
\caption{\label{fig:negdisperse}Squared growth-rate as a function of scaled
wavelength when $X<0$.  The linestyle identifications are the same as in
Fig.~\ref{fig:posdisperse}.}
\end{figure}

\begin{figure}
\centerline{\psfig{file=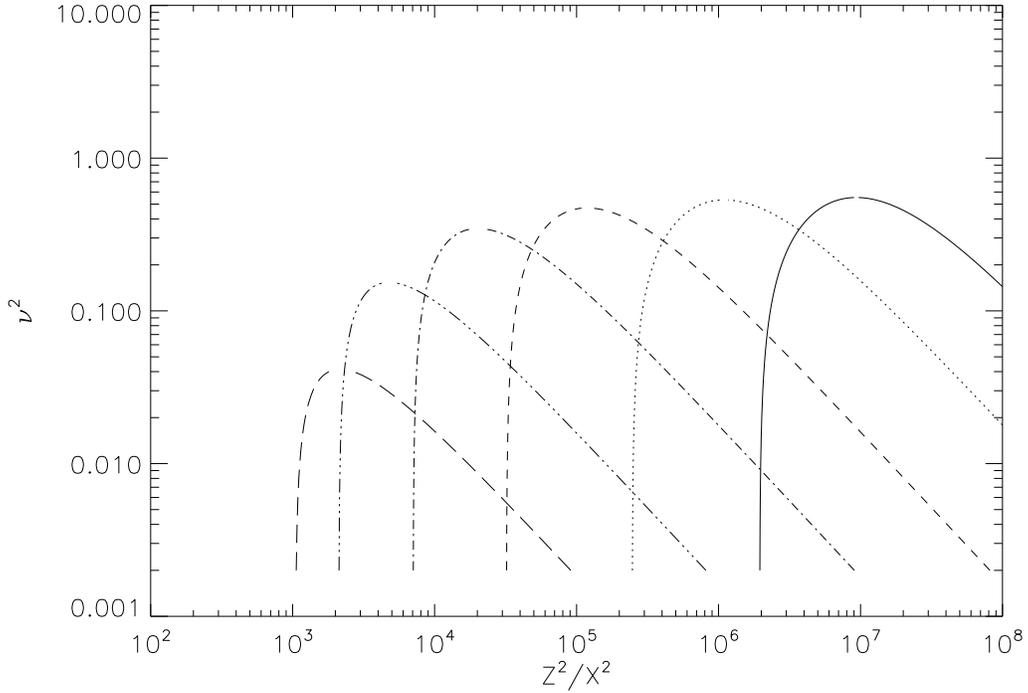,angle=90,width=5.5in}}
\caption{\label{fig:smalldisperse}Squared growth-rate as a function of scaled
wavelength for very small $|X|$.  The linestyle identifications in this figure
are: $X=-10^{-5}$: solid; $X=-3\times 10^{-5}$: dotted; $X=-10^{-4}$:
dashed; $X=-3\times 10^{-4}$: dot-dashed; $X=-10^{-3}$: triple-dot-dashed;
$X=-3\times 10^{-3}$: long-dashed.  To put these in perspective, note that
$1/R = 5.44 \times 10^{-4}$.  Note, too, that the $Z^2/X^2$ scale in this
figure is offset, relative to the scale in the previous two figures, by a factor
of $10^3$.  When $|X|$ is not too small (merely $\sim 10^{-3}$), the
growthrate is still suppressed by Hall effects, even for anti-aligned
field and rotation; at still smaller $|X|$, the dispersion relation
becomes independent of $X$, depending instead only on $RZ^2$.}
\end{figure}

Exact solutions of the full dispersion relation are shown in
Figures~\ref{fig:posdisperse}, \ref{fig:negdisperse}, and \ref{fig:smalldisperse}
in order to show the transitions between these regimes and
confirm our analytic approximations.
As illustrated in Figure~\ref{fig:posdisperse}, when the field and
the rotation rate are aligned, as $X$ falls toward $\sim O(1)$ and
below, the weakening ability of the field to control the ion motion leads
both to a narrowing in the range of unstable wavelengths and to a reduction
in the maximum growthrate.

  When $|X| \sim 1$ but $X<0$, as predicted by our analytic approximation,
there is rapid growth at short wavelengths even while
the same wavelength range is
damped for positive $X$ (Fig.~\ref{fig:negdisperse}).  The reason for this
contrast is that when $X > 0$, the Larmor motion opposes the orbital rotation
where the orbital frequency is greater and adds to the orbital rotation
where it is smaller, whereas it is the other way around
for $X < 0$.  Balbus \& Terquem (2001) and Wardle (1999) similarly found
that disks with field and orbital frequency anti-parallel are more unstable
than those in which they are parallel.  At somewhat smaller values of $|X|$,
the behavior becomes more similar for the two opposite signs.

   Our analytic approximation for $|X| < 1/R$ is also
confirmed by the full numerical solution (Fig.~\ref{fig:smalldisperse}), which
finds a minimum wavelength for growth $\sim R^{-1/2}$, a maximum growthrate
at wavelengths a few times longer, and a growthrate that declines
$\propto Z^{-1}$ for still longer wavelengths.  In this
regime, the field-strength parameter $X$ doesn't appear explicitly
in the dispersion relation, eliminating any dependence on the magnetic
field strength.  In other words, when $|X|$ is this small, the
instability dynamics are no longer magneto-rotational.  They are better
described as the result of current fluctuations that inductively
drive electric field fluctuations with the appropriate phase relative
to the gravitational force as to cause instability.  Because
$|\vec E_1| \sim [\sigma/(kc)^2] |\vec J_1|$, fluctuations on
wavelengths that are too short are ineffective, and there is a minimum
wavelength for instability.   This minimum wavelength must be of order
the characteristic lengthscale, which in this case is the electron
inertial length ($\delta_e = R^{-1/2}\delta_i$).  The coupling to
gravity makes the characteristic timescale $\sim 1/\Omega$, but only on
the short scales $s\sim 1$.

The instability for  $|X| < 1/R$ can be recovered directly from
eqns.~(\ref{e:rmotion}) and (\ref{e:tmotion}) by dropping the terms in
which $\omc$ appears, computing $\mbfj_1$ from the electron motion alone,
and using eqn.~(\ref{e:jE}). The resulting dispersion relation is exactly
eqn.~(\ref{e:vsmallx}). In this limit, the perturbed electron motion is
diagonally across the unperturbed orbit, oriented such that inward radial
motion is in the forward azimuthal direction.  This electron motion
creates a perturbed oscillating electric field of the same orientation
and a perturbed oscillating magnetic field with complementary angle to
the orbit and phase offset by $\pi/2$.  Because the magnetic field
is larger in amplitude than the electric field by a ratio
$\sim \omega_{pe}/\Omega$, the mean electromagnetic stress is dominated
by the magnetic part; it is $\langle -B_{1r}B_{1\phi}\rangle =
(2/\sqrt{3s-1})|B_{1\phi}|^2$, so that it transports angular momentum outward.
It can similarly be shown from eqns. (\ref{e:r1}) and (\ref{e:t1}), together
with the dispersion relation for unstable modes, that the torque on an electron,
$-eE_{1\phi}r_0$, is in phase with the radial displacement $r_1$.
That is, an outwardly (inwardly) displaced electron has a positive
(negative) angular momentum perturbation, which reinforces the motion.
Thus, this mode might fairly be called an
``electromagnetic-rotational instability".

%For example, at $X=10$, the fastest growing mode has $Z=11.8$ and growth
%rate $\nu=0.698$, while for $X=-10$ the fastest growing mode has $Z=8.6$
%and growth rate $\nu=0.818$.  

Summarizing these results, we have found that the most rapidly growing
wavelength has $Z \sim |X|$ or $k\sim\Omega/v_A$ for all field strengths
$|X| \gtrsim 1/R$.   The maximum growth rate $\nu_{\rm max} \sim 1$ when
$|X| \gg 1$, but is smaller when $1/R \lesssim X \lesssim 1$ and
$1/R \lesssim -X \lesssim -3$.  The magnetic character of instability
finally disappears when $|X| < 1/R$, leaving an electromagnetic-gravitational
mode that is unstable for $Z > 0.02$ and grows at a rate $\nu \simeq 1$
on scales of order the electron inertial length $\delta_e$, and at much
slower rates on global scales.

  For all cases in which $|X|$ is not
$\gg 1$, the physical scale of these modes becomes very short compared
to any likely structural scale.  For example, if the plasma is located in
a geometrically-thin disk that is supported by its own gas pressure against
the vertical component of gravity, the wavelength of the most rapidly-growing
mode is a fraction $v_A/c_s$ of the vertical scale-height; when the field is
very weak, this is likely to be an extremely small fraction.
It is not clear whether such small scale disturbances can play a significant
role in either angular momentum transport or dynamo-generation.  They may
be damped by other processes (see \S~\ref{s:damping}); even if they are
not damped, inverse cascades in the nonlinear regime may be necessary for
them to have significant global effect.

\subsection{Long wavelength limit}

Therefore, we devote special effort to understanding the nature of
instability at long wavelengths, even if the modes are slowly growing.
By limiting consideration to large $Z$, we can simplify eqn.~(\ref{e:DR})
to the point that we may derive approximately by analytic means some of
the numerical results just illustrated.

To lowest-order in $Z^{-2}$, eqn.~(\ref{e:DR}) reduces to
\begin{equation}\label{e:highnu}
G_{\nu}^2+\left(\nu^2-3\right)F_{\nu}^2=0.
\end{equation}
The solution in this limit is $\nu = \pm i$, or stable
epicyclic oscillations.  To next lowest-order, we find a low
growthrate unstable mode with
\begin{equation}\label{e:lownu}
\nu^2\approx\frac{3F_0}{Z^2\left(G_0^2-3F_0^2\right)}.
\end{equation}
This solution corresponds to the low growthrate, long wavelength tail
seen in Figures~\ref{fig:posdisperse} and \ref{fig:negdisperse}.
For $Z^2/X^2 \gg 1$, it gives an
approximation to the correct scaling, but overestimates the maximum
growth rate by $\sqrt{5}$.

For a hydrogen plasma, the right hand side of eqn.~(\ref{e:lownu}) is
positive, signifying instability for all $X$. In
the MHD limit, $\nu^2\sim 3X^2/Z^2$, and this approximation holds quite
well even for $X^2\sim 1$. If $RX<1$, $\nu^2$ approaches the limiting value
$3/(Z^2 R)$, independent of $X$.  Thus, although cold plasma theory predicts
that there is no minimum magnetic field necessary for instability, the growth
rate at macroscopically interesting wavelengths is very slow.

\subsection{Equal mass charge-carriers}\label{s:equalmass}

    In certain circumstances (e.g., electron-positron pair plasmas,
dust grains as the predominant charge carriers), the charge/mass ratio
for both the positive and negative charges in the plasma may be the same,
and therefore both have the same magnitude cyclotron frequency.  In that
case, the parameter $R=1$, and the dispersion relation changes form.

\begin{figure}
\centerline{\psfig{file=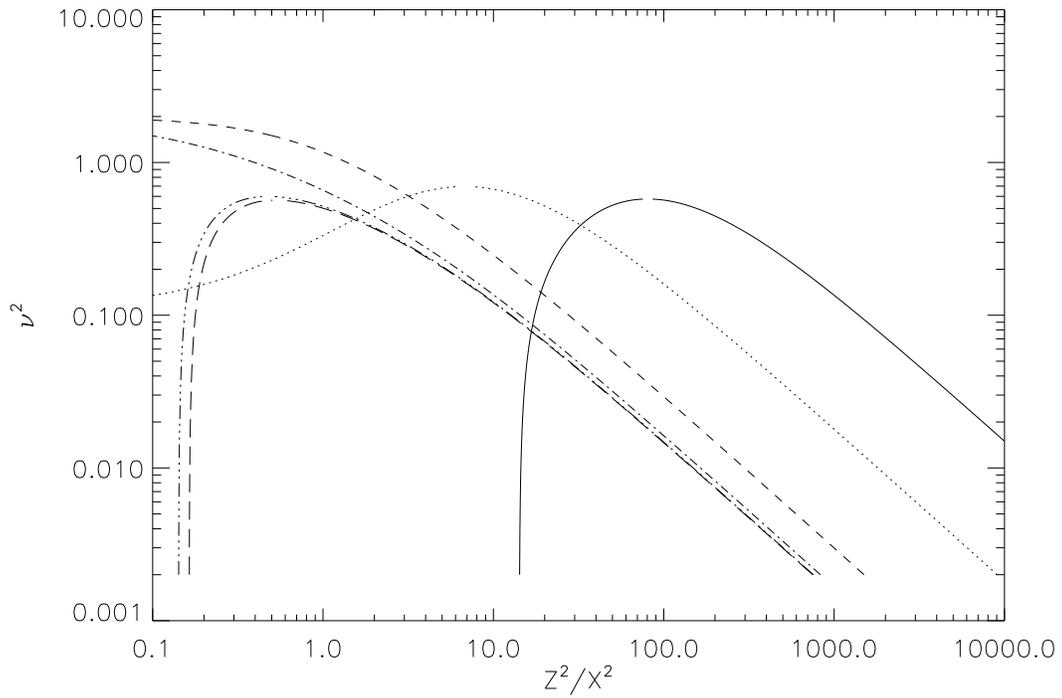,angle=90,width=5.5in}}
\caption{\label{fig:pairdisperse}Squared growth-rate as a function of scaled
wavelength for a plasma in which $R=1$.  For all the curves shown here,
$X>0$, and the linestyle identifications are the same as in
Fig.~\ref{fig:posdisperse}: $|X|=0.1$: solid; $|X|=0.5$:
dotted; $|X|=1.0$: dashed; $|X|=2.0$: dot-dashed; $|X|=10.0$:
triple-dot-dashed; $|X|=100.$: long-dashed.}
\end{figure}

     We find that there are modest, but interesting, changes in the character
of the magneto-rotational instability (see Fig.~\ref{fig:pairdisperse}).
So long as $X$ is greater than a few, the maximum growthrate is $\simeq 1$
and the wavelength of most rapid growth has $Z/X = (kv_A/\Omega)^{-1} \sim 1$.
When $X \sim 1$, growth becomes possible at all wavelengths, not just for
those longer than a critical length.   At still smaller $X$, although the
maximum growthrate remains $\sim O(1)$, there is once again a minimum
wavelength for growth.  Just as in the ordinary plasma case, this
minimum wavelength is fixed at the inertial length; here, however, this
length is the same for the species carrying both signs of charge.
This behavior sets in at $|X| \lesssim 1$ because the ratio $R = 1$.

\section{Mode Structure and Angular Momentum Transport}

Before considering possible damping mechanisms, we evaluate the efficiency
of the MRI in the weak-field regime for driving angular momentum transport
in disks.  We begin by computing the vertically averaged energy density
$\langle W\rangle$ and shear stress $\langle\Pi_{r\phi}\rangle$
associated with an MRI mode:
\begin{equation}\label{e:W}
\langle W\rangle=\frac{\langle B_{1r}^2+B_{1\phi}^2\rangle}{8\pi}+
\frac{1}{2}\rho\left(\langle v_{1r}^2\rangle+\langle v_{1\phi}^2\rangle\right).
\end{equation}
\begin{equation}\label{e:jflux}
\langle\Pi_{r\phi}\rangle=-\frac{\langle B_{1r}B_{1\phi}\rangle}{4\pi}
+\langle\rho v_{1r}v_{1\phi}\rangle,
\end{equation}
where the angle brackets denote vertical 
averages and $\rho$ is the mass density. The ratio
\begin{equation}\label{e:cale}
{\mathcal{E}}\equiv - \frac{\langle\Pi_{r\phi}\rangle}{\langle W\rangle}
\end{equation}
measures the efficiency of angular momentum transport at fixed mode
energy density. The sign is chosen so as to make ${\mathcal{E}}$ positive
for MRI modes.

The ion velocities can be expressed in terms of the magnetic field
fluctuations using Faraday's Law and eqns.
(\ref{e:r1}), (\ref{e:t1}), and (\ref{e:vars}) as
\begin{equation}\label{e:r1b}
v_{1r}=\frac{iq\nu}{mck}\frac{\nu B_{1\phi}-(X+2)B_{1r}}{X^2+4X+1+\nu^2},
\end{equation}
\begin{equation}\label{e:t1b}
v_{1\phi}=-\frac{iq\nu}{mck}\frac{(X+2)B_{1\phi}+\nu(1-3/\nu^2)B_{1r}}{X^2+4X+1+\nu^2}.
\end{equation}

In the MHD limit, the ions follow the fieldlines, and eqns. (\ref{e:r1b}) and (\ref{e:t1b}) reduce to
\begin{equation}\label{e:r1mhd}
v_{1r}\rightarrow -\frac{iq\nu}{mck}\frac{B_{1r}}{X},
\end{equation}
\begin{equation}\label{e:t1mhd}
v_{1\phi}\rightarrow -\frac{iq\nu}{mck}\frac{B_{1\phi}}{X}.
\end{equation}
Substitutings eqns. (\ref{e:r1mhd}) and (\ref{e:t1mhd}) into eqns. (\ref{e:W}) and (\ref{e:jflux}) and
using eqns. (\ref{e:vars}) gives the energy density and shear stress in the MHD limits
\begin{equation}\label{e:Wmhd}
\langle W\rangle\rightarrow  \frac{\langle B_{1r}^2+B_{1\phi}^2\rangle}{8\pi}\left(1+\frac{\nu^2Z^2}{X^2}\right)
\end{equation}
\begin{equation}\label{e:Pimhd}
\langle\Pi_{r\phi}\rangle\rightarrow -\frac{\langle B_{1r}B_{1\phi}\rangle}{4\pi}\left(1+\frac{\nu^2Z^2}{X^2}\right).
\end{equation}
The efficiency parameter for MHD modes is
\begin{equation}\label{e:calemhd}
{\mathcal{E}}\rightarrow\frac{2 B_{1\phi}/B_{1r}}{1+(B_{1\phi}/B_{1r})^2}.
\end{equation}
Equation (\ref{e:calemhd}) shows that in the MHD limit, the efficiency of angular momentum transport is optimized for $\vert
B_{1\phi}/B_{1r}\vert = 1$. It follows from eqns.~(\ref{e:r1}), (\ref{e:t1}),
(\ref{e:F}) and (\ref{e:G}) that this ratio is given by
\begin{equation}\label{e:BrBt}
\frac{B_{1\phi}}{B_{1r}}=\frac{\nu Z^2G_{\nu}}{1+Z^2\nu^2F_{\nu}}.
\end{equation}
In the MHD limit, eqn. (\ref{e:BrBt}) reduces to
\begin{equation}\label{ebrbtmhd}
\frac{B_{1\phi}}{B_{1r}}\rightarrow -\frac{2\nu Z^2/X^2}{1+\nu^2Z^2/X^2}.
\end{equation}
Perturbations at the peak growth rate  ( $X^2/Z^2=\frac{15}{16}$,
$\nu=\frac{3}{4}$) are maximally efficient, with
$\vert B_{1\phi}/B_{1r}\vert ={\mathcal{E}}=1$. 
The fieldline tilt $B_{1\phi}/B_{1r}$ decreases smoothly with increasing
$X$ at constant $Z$. Stable modes (imaginary $\nu$) do not contribute to
transport.  At the peak growth rate, the fluid contributions to the energy
density and torque are 3/5 of the magnetic contributions (Alfv\'en waves,
in contrast, are in exact energy equipartition). The ratio of kinetic
to magnetic stress density and energy density increases
smoothly with decreasing $X$ and fixed wavelength.

We now consider departures from MHD. At large fixed $Z$ but decreasing $X$
(long wavelength modes in a weak magnetic field), the magnetic field becomes
tightly wound: $B_{1\phi}/B_{1r}\rightarrow -\sqrt{3}Z/2X$, and the efficiency
parameter $\mathcal{E}$ becomes low.  Kinetic energy grows to dominate
magnetic energy, approaching an asymptotic value of 3/4 the mode energy.

Angular momentum is, however, transported efficiently in the weak-field limit
by short wavelength modes, with $Z/X\sim 1$.  As we saw in \S~\ref{s:results},
these modes have growth rates of order unity when the field and rotation are
anti-parallel, and grow, albeit with reduced rate, when the field and rotation
are parallel.  The degree of fieldline winding,
efficiency factors, and ratio of kinetic to magnetic energy are all
similar to long wavelength modes with the same $Z/X$.

In the extreme weak field case, $\vert RX\vert < 1$, we found that 
perturbations with $s\equiv RZ^2 > 1/3$ are unstable, and that for $s\gg 1$,
$\nu\approx\sqrt{3/s}$. Because these modes grow slowly and
have $(Z/X)^2 = Rs/(RX)^2\gg 1$, we expect angular momentum
transport to be inefficient.  This expectation is borne out by calculations.
The ions play very little role in the dynamics: the electron kinetic energy
and shear stress are larger than their ion counterparts by a factor of $R$.
The magnetic stress and energy density are even larger: they exceed the electron Reynolds stress
and kinetic energy density 
by factors of $s/6$ and $s$, respectively. And the field is tightly
wound: $B_{1\phi}/B_{1r}\sim \sqrt{3s}/2$.  Consequently, the efficiency
${\mathcal{E}}\sim 4/\sqrt{3s}\ll 1$. 

To summarize, we find that over the entire range $\vert RX\vert < 1$ to
$\vert X\vert\gg 1$, angular momentum transport by long wavelength, slowly
growing modes is rather inefficient.  In the range $\vert X\vert\ge R^{-1}$,
angular momentum transport by long wavelength modes decreases in efficiency
as $B$ decreases, while transport efficiency is roughly constant, and close to
maximal, for fixed $Z/X$, at the wavelength of the maximum growth rate.

\section{Estimated Kinetic Damping and Thermal Effects}\label{s:damping}

Pressure forces and dissipative processes can modify the MRI. Although a
complete treatment of these effects is beyond the scope
of this work, their potential importance warrants a synthesis of
what is already known, together with some simple extrapolations and estimates.

Thermal effects vary with collisionality.  A perturbation of wavenumber $k$ 
is collisionless if the particle mean free path $\lambda_\alpha$
for species $\alpha$ satisfies
$\lambda_\alpha k > 1$.  In a hydrogen plasma, the mean free path to Coulomb
scattering $\lambda_\alpha \sim 10^4 T_{\alpha}^2 n^{-1}$~cm (all units
are in cgs).
The dependence of $\lambda_\alpha$ on $T$ guarantees that at
sufficiently high temperatures, all perturbations are collisionless.

In the MHD regime, thermal pressure has no effect on force balance in the
MRI because the unstable modes are noncompressive, even when the plasma
$\beta\equiv 8\pi P_{gas}/B^2\gg 1$.  In collisionless plasmas, the components
of the pressure tensor parallel and perpendicular to the magnetic field 
evolve independently.  Quataert, Dorland, \& Hammett (2002) investigated
the MRI in this case, but making the usual assumption that $\vert X\vert\gg 1$.
They found that if the background magnetic field has an azimuthal
component $B_{0\phi}$, the fastest growing modes are global and vertically
propagating; $\mbfk=\hat\mbfz k$; $k\sim H^{-1}$. At high
$\beta\equiv 8\pi P/B^2\gg 1$, the growth rates are 2-3 times
larger than in the MHD case. 
The persistence of rapidly growing global modes
even when $v_A/H\ll\Omega$ is due to the anisotropic pressure associated with
magnetic field perturbations in the collisionless regime.  However, these
effects disappear if the magnetic field is so weak that the ions are
unmagnetized, because the ion pressure anisotropy is then
reduced.  We therefore expect that a single-pressure description should be
satisfactory when $|X|$ is small.

\subsection{Collisionless damping}\label{ss:collisionless}

Depending on whether $k\lambda_i$ is $\ll 1$ (the collisional regime) or
$\gg 1$ (collisionless), different mechanisms are responsible for damping.
In the collisionless case, damping occurs when energy is exchanged
through a resonance between the
wave and those particles with velocity such that
the wave is stationary in the particle rest frame. The wave is damped if
the slope of the particle distribution function projected along $\mbfB_0$, 
$\partial f/\partial v_{\parallel}$, is negative. The
damping rate is proportional to $\partial f/\partial v_{\parallel}$,
evaluated at the resonant velocity. 

In a homogeneous, nonrotating plasma the principal resonances are the Landau
and cyclotron resonances. The Landau resonance corresponds to
$v_{\parallel}=-Im(\sigma)/k_{\parallel}$\footnote{In a kinetic treatment,
$\sigma^2$ is complex, so the phase velocity has a real part.}.
Particles exchange energy with the wave through acceleration by the parallel
electric field $E_{1\parallel}$.  Landau damping of magnetosonic modes in
high $\beta$ plasmas has been considered by Foote \& Kulsrud (1979), and
Landau damping of the MRI has been calculated by Sharma, Hammett \& Quataert
(2003). Landau damping of vertically propagating modes is very weak, because
$E_{1\parallel}$ is very small---if $k_r\equiv 0$ it vanishes exactly, and
is nonzero only because of the radial inhomogeneity of the disk. In particular,
the fastest growing modes of ideal theory, $k\sim\Omega/v_A$, which have very
short wavelength in a weak magnetic field, are only weakly damped.

The ion cyclotron resonance damps through energy exchange with the perpendicular
electric field $\mbfE_{1\perp}$. The resonance condition is modified by
gravitational effects to
$$Re\left[\left(\sigma + ik_{\parallel}v_{\parallel}\right)^2 + \omega_{ci}^2 +
    4\Omega\omega_{ci} + \Omega^2\right] = 0,$$
which reduces to the standard expression for
$\vert\Omega/\omega_{ci}\vert = |X|^{-1} \ll 1$.
In a weak magnetic field, resonant particles lie well within the core of
the velocity distribution function, where $\partial f/\partial v_{\parallel}$
is small and cyclotron damping is weak.

Resonant damping of the longer wavelength instabilities presents a somewhat
different case.  In a homogeneous plasma, the damping rates would be larger
than for the short wavelength modes: Landau damping because $E_{1\parallel}$
is relatively larger, and cyclotron damping because the resonant velocity is
larger. However, for global modes the vertical oscillations of the disk
particles must be taken into account: the resonance is thereby weakened and the damping
reduced (Sharp, Berk \& Nielsen 1979, Koepcke et al 1986 \& references therein).
Although none of the work cited applies directly to the MRI, we expect weaker
Landau damping and cyclotron damping of the global modes than in a homogeneous
plasma.  Altogether then, collisionless damping, at least at linear
amplitudes, is probably a weak effect.

\subsection{Collisional damping}

The energy
of a perturbation with vertical wavenumber $k$ is dissipated by viscosity and resistivity
at the rates
\begin{equation}\label{e:Qv}
Q_v=D_vk^2\rho\langle v_1^2\rangle,
\end{equation}
\begin{equation}\label{e:Qr}
Q_r=D_rk^2\frac{\langle B_1^2\rangle}{4\pi},
\end{equation}
where $D_v$ and $D_r$ are the viscous and resistive diffusivities, respectively.
The viscous diffusivity is related to the ion thermal velocity $v_i$, collision
time $\tau_i$, and mean free path $\lambda_i$ by
$D_v \sim v_i\lambda_i\sim v_i^2\tau_i$. The resistive diffusivity is related
to the electron inertial length $\delta_e$ and collision time $\tau_e$ by
$D_r\sim\delta_e^2/\tau_e$.  Equation (\ref{e:Qv}) is valid
for wavelengths longer than $\lambda_i$. Equation (\ref{e:Qr}) requires lengthscales
longer than both $\lambda_e$ and $\delta_e$ in order for the form of Ohm's law on which it is
based to be valid.

If there were no unstable growth, the collisional damping rate
$\Gamma_c$ could be read off from the wave energy equation
\begin{equation}\label{e:Edot}
\frac{dW}{dt}=-2\Gamma_c W = Q_v+Q_r.
\end{equation}
%
%In eqn.~(\ref{e:Edot}) it is assumed that growth and damping mechanisms are
%additive.
Substituting eqns.~(\ref{e:Qv}) and (\ref{e:Qr}) into
eqn.~(\ref{e:Edot}), defining the magnetic Prandtl number $P_{rm}\equiv D_v/D_r$,
and denoting by $f_m$ the fraction of mode energy in magnetic form gives
\begin{equation}\label{e:gammac}
\Gamma_c=k^2D_v\left[f_m P_{rm}^{-1}+(1-f_m)\right]\equiv k^2D_{eff}.
\end{equation}
The $k^2$ dependence of $\Gamma_c$ is characteristic of diffusive damping.
The effective diffusion coefficient $D_{eff}$ is a combination of the viscous
and magnetic diffusivities, weighted by the fractions of kinetic and
magnetic energy in each mode. The relative importance of viscous and
resistive damping is set by $P_{rm}$.
In a hydrogen plasma, $P_{rm}=(\lambda_i^2/\delta_e)^2\tau_e/\tau_i\sim 10^{-5}
T_i^{5/2}T_e^{3/2}n^{-1}$. 
Perturbations in hot, low density disks are damped primarily by viscosity,
while perturbations in cold, dense disks are
damped primarily by resistivity.

Equation (\ref{e:gammac}) is valid
only at wavelengths long enough that eqns. (\ref{e:Qv}) and (\ref{e:Qr})
apply. When $k\lambda_i > 1$, viscous damping saturates at the rate
$\Gamma_{vs}\sim (1-f_m)/\tau_i$.
Resistive damping saturates when $k^{-1}< \max{(\delta_e,\lambda_e)}$.
If $\delta_e >\lambda_e$, the saturated resistive damping rate
$\Gamma_{rs}\sim f_m/\tau_e$, while if $\delta_e <\lambda_e$, 
$\Gamma_{rs}\sim (\delta_e /\lambda_e)^2 f_m/\tau_e$. In subsequent discussions,
we adopt for simplicity the saturated damping rate
\begin{equation}\label{e:gammacs}
\Gamma_{cs} = \frac{1-f_m}{\tau_i}+\frac{\delta_e^2}{\delta_e^2+\lambda_e^2}\frac{f_m}{\tau_e}.
\end{equation}
Equation (\ref{e:gammacs}) applies when $k \lambda_i > 1$ and
$k \min(\delta_e, \lambda_e) > 1$.

If the ions are magnetized ($X\gg 1$) but collisionless, viscous momentum
transport across magnetic fieldlines is drastically reduced (Braginskii
1965). Balbus (2004) has shown that the resulting anisotropic stresses
are destabilizing, much as Quataert, Dorland, \& Hammett (2002) showed
that anisotropic pressure is destabilizing. Diffusive damping is still
present, but is reduced by a factor of $\omc\tau_i$.

\subsection{Consequences}

In order to pull together these ideas, we perform a thought experiment in
which $B$ is increased from an arbitrarily small value and follow the damping
as a function of fieldstrength.

Suppose that at first $\vert X\vert\ll 1/R$. The modes are then sustained
by electrons only, the ions having dropped out (see eqn. \ref{e:vsmallx}). 
Therefore we take $f_m=1$ in eqns. (\ref{e:gammac}) and (\ref{e:gammacs}).
At the scale $\delta_e^{-1}$ of the fastest growing mode, resistive damping
is saturated. According to eqn.~(\ref{e:gammacs}), this mode can grow only
if $\Omega\tau_e(1+\lambda_e^2/\delta_e^2) > 1$.  This criterion is independent
of $B$ as long as $\vert X\vert\ll 1/R$. However, we admit that it is based upon
extrapolation, and is not a rigorous result.

Moving up in fieldstrength, we consider  $\vert X\vert \ge 1$, but not too
large. Now both the electrons and the ions participate, and the fastest growing
mode has $k\sim\Omega/v_A\equiv k_{max}$. Because $\vert X\vert\ge 1$,
$k_{max}\delta_i\le 1$, and resistive damping is saturated. Whether viscous
damping is saturated as well depends on $\lambda_i$.

At sufficiently long wavelengths, the MRI growth rate always exceeds the
collisional damping rate, because the growth rate is proportional to $k$ and
the damping rate to $k^2$. Approximating the growth rate by
$\sqrt{3}k v_A$ and using eqn. (\ref{e:gammac}) for the damping rate, we
find that the growth rate exceeds damping at wavelengths long enough that
$k/k_{max}\lesssim\sqrt{3}v_A^2/(2\Omega D_{eff})$.  The growth rate is
approximately $v_A^2/D_{eff}$ at the short wavelength end of this range.
This growth rate is generally a small fraction of the orbital frequency;
quantitatively,
\begin{equation}\label{e:longgrowth}
\nu\sim\frac{v_A^2}{\Omega D_{eff}}\sim\left(\frac{v_A}{v_i}\right)^2\frac{D_v}{D_{eff}}\left(
\Omega\tau_i\right)^{-1}.
\end{equation}
In some circumstances, there can also be growth for $k \sim k_{max}$
because the damping saturates as a function of $k$ at very short wavelengths,
where as the growthrate continues to climb $\propto k$ for all
$k \lesssim k_{max}$.  What determines whether there can be growth at
short wavelength is the ratio
of the maximum ideal growth rate ($\sim\Omega$) to the saturated damping rate,
$\Gamma_{cs}$. If $\Omega/\Gamma_{cs}$ exceeds unity, these modes can grow;
otherwise, they are suppressed.  Because of the form of eqn. (\ref{e:gammacs}),
this depends on the collisionality in the disk. If $\delta_e/\lambda_e\ll 1$,
$\Gamma_{cs}\sim 1/\tau_i$, while for $\delta_e/\lambda_e\gg 1$,
$\Gamma_{cs}\sim 1/\tau_e$.  Thus, in these two limits the criterion for
growth near $k=k_{max}$ is $\Omega\tau_i > 1$ and $\Omega\tau_e > 1$.
Short wavelength growth is therefore
more likely to be found in hot, low density disks with a fast rotation
rate.

    Otherwise, for $\vert X\vert\sim 1$, the MRI is limited to long
wavelength, slowly growing modes as described by eqn. (\ref{e:longgrowth}).
Even these modes may be suppressed, however, by collisionless damping, although we argued in 
\S\ref{ss:collisionless} that collisionless processes are too weak to
stabilize the fastest growing, short wavelength modes. 
An exception to these remarks occurs  in the window
$-2-\sqrt{3} < X<  -2+\sqrt{3}$ (see eqn.~\ref{e:postMHD}). In this range,
$\sigma$ approaches a finite limit, not too different from $\Omega$, as
$k/k_{max}\rightarrow 0$. These modes are only weakly damped by collisional 
effects, permitting rapid growth at large scales.

The picture outlined above is also valid for $\vert X\vert\gg 1$, except that
as the wavelength of the fastest growing mode increases, a much larger
diffusivity is required to suppress the MRI. If $\vert X\vert\gg 1$, 
$\omci\tau_i\gg 1$, and $\mbfB$ has even a small  component in the plane
of the disk, stresses associated with anisotropic pressure and
viscosity can destabilize the disk (Quataert, Dorland, \& Hammett 2002,
Balbus 2004). In a hydrostatic disk, the value of  $\vert X\vert$ at which
these effects come into play is of order $H/\lambda_i$, which is large
except in very hot, low density disks.  For this reason, these effects
are generally unimportant in the weak-field regime, where
$|X|$ is not very large.

\section{Implications}\label{s:implications}

The analysis presented here is applicable to any situation in which the
Larmor frequency of a major charged species is not much larger than the
dynamical frequency. It can also be used to analyze the MRI in media
such as pair plasmas or charged-grain fluids in which the dominant positive
and negative species have equal mass. The primary applications we have in
mind involve situations where the field is weak. 

How weak is weak? At the order of magnitude level, it is always possible
to estimate the
dynamical frequency of orbital motion in terms of the smoothed-out density
of mass contained within the orbit: $\Omega \sim \sqrt{G \overline{\rho}}$.
Using that estimate, our field-strength parameter $X\equiv\omega_{ci}/\Omega$
may be rewritten in an instructive way:
\begin{equation}\label{e:alfven}
X = \frac{e}{m_i c}\sqrt{\frac{4\pi}{G}} \overline{v}_A = 1.3 \times 10^8
     \left(\overline{v}_A/\hbox{~cm$^{-1}$~s}\right),
\end{equation}
where $\overline{v}_A$ represents the Alfven speed estimated using the
smoothed-out density.  When the bulk of the gravitating mass is distributed
and has only modest density gradients, $\overline{v}_A$ is a fair estimate
of the actual local Alfven speed; when the gravitating mass is concentrated,
$\overline{v}_A$ is only an indicator of a characteristic speed.  In either
case, it is clear from the numerical value of the coefficient in
equation~\ref{e:alfven} that the characteristic Alfven speed must be
truly tiny---$\lesssim 10^{-8}$~cm~s$^{-1}$---for $X \lesssim 1$.

Before beginning to explore some applications of these ideas, we briefly
summarize our results so far.  We have shown that an ideal plasma with a
magnetic field oriented along the rotation axis is always unstable,
and there is almost always a mode that grows at a rate comparable to the rotation
frequency $\Omega$ (the exception is the range $1/R \lesssim X \lesssim 1$,
in which the maximum growth rate is somewhat slower when the field and the
angular velocity are in the same direction).  When
$\vert X\vert\ll m_e/m_i\equiv R^{-1}$, the fastest growing mode has a
wavelength on the scale of the electron skin depth
$\delta_e\equiv c\sqrt{m_e/(4\pi n_ee^2)}\sim 5n_e^{-1/2}$km.  The smallness
of this scale and the low efficiency at which these
modes transport angular momentum leaves their significance open to question.

For $\vert X\vert\sim 1$, the instability is sensitive to whether $\mbfB$ and
$\Omega$ are parallel or antiparallel, with larger growth rates in the
latter case.  This asymmetry is a manifestation of the different
charge/mass ratios for charge carriers of opposite sign: the lower-mass
charge-carrier remains well-magnetized even while the opposite sign
charge-carrier, of opposite charge sign, deviates from the fieldlines.
Except for the range  $-2-\sqrt{3} < X<  -2+\sqrt{3}$, in which growth rates of
order $\Omega$ persist at arbitrarily long wavelengths, the fastest growing
mode is near the wavenumber $k_{max}\equiv\Omega/v_A$, and the growth rate
scales linearly with $k$ for $k<k_{max}$.  This behavior is similar to what
is predicted by ideal MHD, even though $\omega_{ci}$ need not be much greater
than the dynamical frequency.  The fastest growing modes are also the most
efficient at transporting angular momentum.

Thermal effects modify this picture. If $\mbfB$ has an azimuthal component
and the ions are magnetized ($\vert X\vert\gg 1$) but have a long mean free
path, there are modes with $k\ll k_{max}$ with growth rates near $\Omega$
(Quataert, Dorland, \& Hammett 2002, Balbus 2004).  These effects do not appear if $\mbfB$
is vertical, but thermal effects are still substantial in that they cause
damping.  Although collisionless damping appears to be weak, especially for the short
wavelength modes, collisional
damping---both viscous and resistive---can be strong. It increases with
wavenumber as $k^2$ until it saturates at the ion mean free
path $\lambda_i$ (viscous damping) or $\max{(\lambda_e,\delta_e)}$
(resistive damping). Although growth dominates damping at sufficiently long
wavelengths, the MRI is suppressed at $k\sim k_{max}$ unless the maximum growth 
rate, $\Omega$, exceeds the saturated damping rate
$\Gamma_{cs}$.  This criterion is most easily fulfilled in hot, low
density disks in which the particle mean free paths and skin depths are long.
Otherwise, the growth rate of the MRI peaks at the rather low value given by
eqn.~(\ref{e:longgrowth}). Even this is an upper limit because
it neglects collisionless damping. 

These results apply to a number of situations in which a seed magnetic field
has been generated at early cosmological epochs.  For definiteness we suppose
the mechanism of magnetogenesis is the Biermann battery.  When that is the
case, there is a characteristic relation between the operation of the battery
and possible MRI growth.  After developing this relationship, we 
briefly address two questions: at what fieldstrength can angular momentum 
transport by the
MRI begin to play a role in disk accretion, and at what
scales can the MRI generate strong
magnetic fields? 

\subsection{Interplay with the Biermann battery}

The ``Biermann battery" (Biermann \& Schl\"uter 1951) 
depends on misaligned gradients in the electron density and
temperature.  Field generation by this mechanism occurs at a rate
\begin{equation}\label{e:biermann}
\frac{\partial\mbfB}{\partial t}=-\frac{ck_B}{en_e}\mbfnabla n_e\times
\mbfnabla T_e,
\end{equation}
where $k_B$ is Boltzmann's constant. 
If we define $l_n$ and $l_T$ as the characteristic scales for the density
and temperature gradients, respectively, and suppose that there is a sizable
angle between the two gradients, the characteristic field amplification time
$t_{\rm Bier}$ equired to reach a particular value of $X$
may be estimated in dimensionless form as
\begin{equation}\label{e:biermannest1}
\Omega t_{\rm Bier} \sim X \Omega^2 l_n l_T v_{*}^{-2},
\end{equation}
where $v_* \equiv (kT_e/m_i)^{1/2}$ is thermal speed of the ions if
$T_i = T_e$.  Not surprisingly, the relative importance of the Biermann
mechanism grows with diminishing $X$.  Defining $H_* \sim v_*/\Omega$,
we find that
\begin{equation}\label{e:biermannest}
\Omega t_{\rm Bier} \sim X l_n l_T/H_*^2.
\end{equation}
In a conventional radiative accretion disk, $H_*$ is, of course, the
disk thickness $H$ because $T_e = T_i$.  In the absence of collisional
damping, the characteristic growth time of
the MRI is $\sim\Omega^{-1}$, so field amplification by the Biermann battery
would outstrip the MRI until the time when the right hand side of
eqn.~(\ref{e:biermannest}) is greater than unity.  If
the gradient scales are comparable to the characteristic thermal travel
distance, the Biermann battery field augmentation rate surpasses
that of the MRI whenever $|X| < 1$ and even to somewhat greater values
of $X$ when $X > 0$.  However, if one of the gradient scales is
${\mathcal{O}}(r)$, the MRI would overtake the battery somewhat sooner,
when  $|X|>H_*/r$ (particularly if $X < 0$).

On the other hand, collisional damping can in many circumstances substantially
reduce the growth rate (e.g., as in the case described by
eqn.~\ref{e:longgrowth}).  When that is the case, the Biermann battery
dominates as long as
\begin{equation}\label{e:biermannMRI}
|X|<\left(\frac{T_i}{T_e}\frac{D_{eff}}{D_v}\frac{H_*^4}{l_nl_T\delta_i^2}\Omega\tau_i\right)^{1/3}.
\end{equation}
The right hand side of eqn. (\ref{e:biermannMRI}) can be much greater than one,
because of the smallness of $\delta_i$ compared to global scales.  Here longer
collision time leads to a wider range of battery dominance because it also
means that the minimum wavelength for growth is greater, and that suppresses
the MRI growth rate.

The Biermann battery can also operate in shock fronts and ionization
fronts propagating in structured media. These contexts can be of
interest in cosmological and galactic settings (Kulsrud et al. 1997,
Gnedin, Ferrara, \& Zweibel 2000). One of the lengthscales $l_n$ or $l_T$
is replaced by the thickness of the front, $l_f$, but the battery operates
within each fluid element for only as long as it takes the front to
cross it, $l_f/v_f$. The resulting value of $\omega_{ci}$ is
\begin{equation}\label{e:omegacf}
\omega_{cf}\sim\frac{v_*^2}{lv_f},
\end{equation}
where $l$ is the (global) lengthscale on which $n_e$ or $T_e$ varies. In the relevant settings
we expect $T_i\sim T_e$, in which case eqn. (\ref{e:omegacf})
shows that the ion cyclotron radius $r_i\sim v_*/\omega_{cf}$ is of order $l$, the
global gradient scale. Other plasma properties being equal, the 
fields produced in disks exceed the fields produced in fronts after a time
of order $l/v_f$.

We now apply these generic results about the complementary relationship
between Biermann batteries and the MRI in some specific settings.  At
least in these examples, we find that the MRI has little impact because
collisional damping is so strong when the field is extremely weak.

\subsection{Cosmological scales}

One particular setting in which these effects may act is the initial growth
of density perturbations at high redshift.
It is possible that the only magnetic field at that point is the field
arising from the Biermann battery.
Starting from negligible field, after one Hubble time we might
expect
\begin{equation}
X \sim \frac{kT_e}{m_i H_0^2 l_{n0} l_{T0}} \sim 0.02 (1+z)^{-1} T_{e4}
          l_{n0,Mpc}^{-1} l_{T0,Mpc}^{-1},
\end{equation}
where $H_0$ is the present-day Hubble constant, $l_{n0}$ and $l_{T0}$
are the co-moving lengthscales of the density
and temperature gradients, and they have been scaled to Mpc for the numerical
estimate.  Thus, under these conditions, any
possible MRI effects would in fact be in the weak-field regime.  Only
after the gas temperature has risen considerably (e.g., due to shocks
in collapsing clouds) and the relevant gradient scales have become
more characteristic of galaxies than of intergalactic gas would the
MRI operate in the ordinary regime.

At the earliest stages of magnetic field amplification, where $|X| < 1/R$,
there could be growth over all possible wavelengths in the electron-dominated
modes provided it takes place at high enough redshift:
$z \gtrsim 200 T_{e4}^{1/2}$.  However, as soon as the modes shift over
to the conventional MRI, damping restricts the range of growing wavelengths
to only those on roughly cosmological scales:
\begin{equation}
k < k_{\rm crit} \sim 10^{-33} (1+z)^2 T_{e4} T_{i4}^{-5/2}
      l_{n0,Mpc}^{-1} l_{T0,Mpc}^{-1}(D_v/D_{eff})\hbox{~cm$^{-1}$}.
\end{equation}
Moreover, the fastest growing electron modes are probably damped by resistivity. 
Thus, this appears to be an unpromising venue for significant MRI effects.

The magnetic fields generated by the Biermann battery operating in cosmological shocks and
ionization fronts are of order 
$10^{-18}$G, corresponding to $\omega_{ci}=10^{-14}$ s$^{-1}$, or less. 
If we assume the velocity shear is such that $|X|\sim 1$ and $n\sim 10^{-3}
$, an appropriate intergalactic density at $z\sim 10$, then $k_{max}\sim 10^{-9}$ cm$^{-1}$.
On the other hand, $\tau_i\sim 10^{9}T_{i4}^{3/2}$, so collisional damping has surely saturated
and $\Omega\tau_i\ll 1$. The MRI cannot grow at these scales.  On the larger scales at which it
is excited,
the growth time exceeds the age of the universe. The MRI appears to be unimportant in the IGM at
large at all times.

\subsection{Galactic scales}

The fully ionized gas in galactic disks has high magnetic Prandtl number
$P_{rm}$, meaning that viscous damping dominates
resistive damping, and due to the long galactic rotation period is
highly collisional---$\Omega\tau_i \ll 1$. Therefore eqn.~(\ref{e:longgrowth})
applies, and the growth rate of the fastest growing
mode is of order $v_A^2/D_v\sim (v_A/v_i)^2/\tau_i\sim
B^2/(4\pi kT^{5/2})\sim 6\times 10^4B^2/T_4^{3/2}$~s$^{-1}$.  This becomes
comparable to the galactic rotation rate only for $B\ge 10^{-10}$~G.  Such
a strong field cannot reasonably be generated by a Biermann battery on
galactic scales.  Additional constraints apply to weakly ionized galactic gas,
in which the MRI is damped by ion-neutral friction.  These results suggest
that the MRI cannot be the primary agent in amplifying
seed magnetic fields generated by the Biermann battery in galactic disks.

\subsection{Disks surrounding massive objects}

The pivotal role of MRI turbulence in accretion motivates this application.
Because, as we have just stressed, so little is known about the cosmological
history of magnetic field development,
it is an interesting question whether the matter accreting onto proto-stars
in the first galaxies or black holes in high-redshift quasars contains
magnetic fields of the magnitude one might expect in contemporary interstellar
gas.  If the seed field in accreting matter then were much weaker than it
would be today, the question arises as to whether the MRI would successfully
amplify the field to the level at which accretion becomes efficient.  Indeed,
there may be a ``bootstrap" issue here: it has been suggested that magnetized
outflows from early disks could have injected the first significant magnetic
field into large portions of the intergalactic medium (Daly \& Loeb 1990,
Kronberg et al. 1999, Furlanetto \& Loeb 2001).  If disks are required to create
magnetic fields, what magnetic fields are present in the first disks that
allow them to function?

For a semi-quantitative approach to answering this question, consider
conditions at the innermost stable circular orbit (ISCO) of a nonrotating
black hole of mass $M$.  Using $\Omega_{ISCO}=c/(6r_g)$ and assuming a
hydrogen plasma, 
eqn.~(\ref{e:vars}) gives
\begin{equation}\label{e:xisco}
X_{ISCO}
=0.3B\frac{M}{M_{\odot}},
\end{equation}
where $B$ is given in G. Thus, the lower the mass of the compact object, 
the stronger the field must be to meet the conditions of MHD.  This is,
of course, a prime example of where eqn. (\ref{e:alfven}) is {\it not} a
good estimator of the relationship between the Alfven speed in the plasma
and the $X$ parameter.  Around
a non-rotating black hole, $X(r)\sim X_{ISCO}(6r_g/r)^{3/2}$.

To put this in perspective, we turn the problem around and ask what value of
$X$ would be expected in an efficiently-accreting disk.  It is very large:
\begin{equation}
X_{\rm eff} \sim 10^{9} \dot m^{7/16} (M/M_{\odot})^{9/16} (r/r_g)^{3/16}
                   \tau^{-1/16},
\end{equation}
where $\dot m$ is the accretion rate in Eddington units and $\tau$ is
the optical depth through the disk.  This estimate, by the way, is
independent of whether the vertical support of the disk is dominated
by gas or by radiation pressure; it assumes only that angular momentum
and energy are conserved, dissipated energy is radiated locally, the
disk is in a steady-state, and that the disk is optically thick so that
the temperature in its interior is $\sim \tau^{1/4}$ times larger than
at its surface.   Thus, significant accretion rates (as measured in
Eddington units) go hand-in-hand with large $|X|$ unless the stress
per unit surface density is so low that the disk accumulates a very large
mass.

Suppose, then, that a disk is formed, but without any magnetic field, and
therefore without any turbulence or inflow.  Can it ever work its way
to a state of efficient accretion?  In order to avoid a singular state,
we must assume that there is some mechanism (other than local dissipation
associated with accretion) that heats it.  Under almost any circumstances,
we expect the disk to be stratified both radially and vertically, and, because
of local heating and cooling processes, to be nonbarytropic.  This, too,
is a context in which the Biermann battery should operate.  

Here the crossed gradients must be in the radial and vertical directions,
so $l_n l_T \sim rH$.\footnote{According to eqn.~(\ref{e:biermann}), vertical and
radial gradients generate an azimuthal field. We assume our analysis of the
MRI for a vertical field applies to this case.}  Making use of
eqn. (\ref{e:biermannest1}) yields
\begin{equation}\label{e:xdot}
\frac{dX}{dt}\sim\frac{v_*^2t}{rH}.
\end{equation}
Explicitly allowing for a possible contrast in ion and electron temperatures
and integrating, we find that
\begin{equation}\label{e:x}
X=\frac{\Omega t}{1+T_e/T_i}\left(\frac{H}{r}\right).
\end{equation}
Thus, the ratio of the growth rate due to the battery to that due to
MRI growth is $\sim \nu^{-1} (h/r)X^{-1}$.  

In the limit of extremely small $|X|$, the preceding expression shows that
magnetic field growth due to the Biermann battery effect is likely to outpace
that due to local instability.   To the degree that $\nu < 1$, due either to
damping at short wavelength or the slower growth rate found at long wavelengths,
this conclusion is strengthened.  The time-dependence of the fluctuation
energy density $\delta E$ in a large-scale mode with $kH \sim 1$ may then be found
by combining eqns.~(\ref{e:lownu}) and (\ref{e:x}):
\begin{equation}\label{e:growth}
\frac{d\delta E}{d\Omega t}=2\nu\delta E\sim\frac{ 2\sqrt{3}}{Z}
\frac{\Omega t}{1+T_e/T_i}\left(\frac{H}{r}\right)\delta E.
\end{equation}
Integrating eqn.~(\ref{e:growth}) over time, we find that for a fluctuation
of initial energy $\delta E_0$
\begin{equation}\label{e:amplification}
\delta E = \delta E_0 \exp{
\frac{\sqrt{3}(\Omega t)^2}{1+T_e/T_i} \frac{H}{r} }.
\end{equation}
According to eqn.~(\ref{e:amplification}), the fluctuations grow by a factor
of $e$ in the normalized time $\Omega t_*$
\begin{equation}\label{e:timescale}
\Omega t_*=\left[\frac{r}{H}\frac{Z}{\sqrt{3}}\left(1+T_e/T_i\right)\right]^{1/2}
\sim (kH)^{-1} (r/\delta_i)^{1/2}(1 + T_e/T_i)^{1/2}.
\end{equation}
Equation~(\ref{e:timescale}) corresponds to a very long amplification time,
except for modes very short compared to a disk thickness.  The reason is the
extremely large ratio $r/\delta_i$.   Little MRI growth can be expected,
therefore, on scales anywhere approaching the disk thickness.

If the disk is turbulent, for example, due to gravitational instability,
the battery operates somewhat differently. A random magnetic field is
generated, and the gradient lengthscales in eqn.~(\ref{e:biermann}) are
the characteristic turbulent scales. The operation of the MRI in such a 
field is beyond the scope of this paper.

\subsection{Disks surrounding stellar mass objects}\label{sec:grains}

Our last example is neither cosmological nor involves the Biermann battery.
In this case, we consider how the MRI operates
when most of the charge in the medium is carried by dust. 
In many circumstances of high gas density and low temperature such
as dense molecular clouds or proto-stellar accretion disks, it
is possible that most of the charge in the plasma may actually reside on
grains rather than ions or electrons (Draine \& Sutin 1987, Nishi et al.
1991).  Because the ability of grains to hold charge depends strongly
on their size distribution, there are substantial uncertainties regarding
the specific parameters of the regimes where this condition may obtain.
Nonetheless, it is of interest to examine the consequences for the MRI
wherever this state may exist.

     The most direct impact is that the charged species traversing Larmor
orbits have masses greater than a proton by a factor $\sim 10^9$ or
more. Consequently, the effective cyclotron frequencies are dramatically
depressed.  Posed in terms of the characteristic Alfven speed for a
self-gravitating system (eqn.~\ref{e:alfven}),
$X \sim 0.1Z_{gr}(m_{gr}/10^9 m_p)^{-1}(\overline{v}_A$/cm$^{-1})$~s
for grain mass $m_{gr}$ and grain charge in electron units $Z_{gr}$.
In the context
of proto-stellar disks, it is natural to compare the grain Larmor
frequency to the expected scale of orbital frequencies:
$X \sim 3 \beta^{-1/2} p_{12}^{1/2}$, where we have
assumed an orbital period of 1~yr and the pressure in the disk is
scaled to $10^{12}$~K~cm$^{-3}$.

     However, our simple two-fluid theory cannot be directly applied to this
situation because it assumes the charge carriers account for the bulk
of the inertia, an assumption that fails badly when grains are the
charge carriers.  An appropriate theory would need to incorporate
a third fluid: the neutral background, including its inertia and the
rate of momentum exchange between the neutrals and both kinds of
charge carriers.  These latter effects were the
primary focus of the papers by Balbus \& Terquem (2001) and Salmeron
\& Wardle (2003).  In fact, in the limit they considered, the cross-field
current was determined by balancing the $\mbfj\times \mbfB$ force
with the ion-neutral drag force.  On the other hand, the estimates
of the previous paragraph demonstrate that there may be contexts in
which linear theories of the MRI that take proper account of resistivity,
Hall effects, and so on, and yet ignore the dynamical effects of a small
ratio of cyclotron frequency to orbital frequency may be equally
incomplete.

      Pending a complete investigation of this problem, we offer a
few rough estimates.  First, we have already noted that our parameter
$X = 2x^{-1}(\mu_e/m_i)$, where $x$ is the Hall parameter defined by
Balbus \& Terquem (2001) and $\mu_e/m_i$ is the ratio of the mass per
electron to the ion mass.  When the plasma is strongly-ionized, Hall
effects go hand-in-hand with weak-field effects; in a state
of weak ionization, Hall effects can be important even when $X \gg 1$.
However, if grains are the charge-carriers, the subscripts ``e" and ``i"
refer to the negative and positive charge-carrying grains, so that
$\mu_e/m_i \sim 1$ and the regime of Hall effects can, as in the
strongly-ionized case, be identified with the regime of weak-field effects.

      Next, let the effective collision frequency between
charged grains and the background neutral medium be $\nu_c$ (i.e.,
this is the neutral-grain collision rate divided by the grain/neutral
mass ratio).  Following Balbus \& Terquem (2001), we estimate the
relative importance of inductive and Ohmic effects by the magnetic
Reynolds number, which may be written in our notation as
\begin{equation}
Re_M \sim X^2 \frac{\rho_{gr}}{\rho} \frac{\Omega/\nu_c}{kv_A/\Omega},
\end{equation}
where $\rho_{gr}/\rho$ is the ratio of (charged) grain mass to neutral mass
and must be $\lesssim 10^{-2}$.  In this expression we have taken the
relevant length scale to be $k^{-1}$.  Given that $\nu_c > \Omega$ in most
cases of this sort, small $X$ effects will also typically be associated
with small magnetic Reynolds number, and therefore are likely to occur in
the context of strongly damped modes, except for $k \ll \Omega/v_A$.  

\section{Summary and Conclusions}

    We have shown that generalizing from cold single-fluid theory to
cold two-fluid theory permits at least an approximate description of
how the magneto-rotational instability behaves as the field becomes
extremely weak.  ``Weak" in this sense may be parameterized in terms
of the ratio between the ion cyclotron frequency and the rotational
frequency, the quantity we have labelled $X$.  Little in the basic
nature of the instability changes so long as $|X| > 1$, although the
characteristic wavelength scale is proportional to $X$, and can therefore
become very short compared to other interesting scales.

    Complications arise when $1 > |X| > 1/R$.  ``Hall effects",
dynamical distinctions between the behavior of the two signs of
charge, become important.  When the field and
the rotation are aligned, so $X$ is positive, the maximum growth
rate for $X$ in that range can be significantly diminished.  On
the other hand, when the field and the rotation are anti-parallel,
giving negative $X$, the maximum growth rate, while reduced somewhat,
remains closer to $\Omega$.
In addition, unstable growth becomes possible at arbitrarily short
wavelengths, even though it is not possible in stronger fields.

     The entire character of the modes changes when $|X|$ falls
below $1/R$, the mass ratio between the negative and positive
charge carriers, and is better described as ``electromagnetic-rotational"
than ``magnetic-rotational".  In this regime, only the lighter
(usually negative)
species participates.  As a result, the characteristic wavelength
changes from $\sim v_A/\Omega$ to $\delta_e$, the electron inertial
length.  Just as for the ordinary MRI, there is a minimum wavelength
for growth that is comparable to the characteristic wavelength, and
the growth rate scales $\propto k$ for longer wavelengths.

     Although the cold fluid approach demonstrates that rapid growth
persists to arbitrarily weak field, it also entails a shift of focus
to progressively shorter wavelengths.   These are exactly the scales
on which thermal effects, not included in the simple fluid picture,
might intervene.  A complete account of these effects is beyond the
scope of this paper, but simple estimates suggest that, as expected,
they are often able to damp all but the longest-wavelength modes.
Because the growth rate typically scales $\propto k$ for wavelengths
longer than the characteristic wavelength, growth can be very slow
for even the fastest modes that escape collisional damping.  Thus,
the ability of the MRI to amplify magnetic field may be severely
weakened even for values of $|X|$ substantially greater than
$\sim 1$.  In other words, to answer one of the questions with which
we began, the magnetic field becomes too weak to be effective when
the preferred scale of instability ($v_A/\Omega$) is so small
that transport-associated dissipation (resistivity and viscosity)
limits growth.

     Lastly, we have applied these ideas to a few astrophysical contexts
in which extremely weak fields might be expected.   On the basis of these
estimates, we find that it may, for example, be very difficult for accretion
disks formed from primordial gas in which no magnetic field has been
embedded to ``bootstrap" themselves from a state of initially weak field to
a state in which the field is strong enough to drive rapid accretion. Likewise, the
MRI is unlikely to be an efficient mechanism for amplifying battery-generated magnetic fields in
the intergalactic medium, or in protogalactic disks; dynamos in these systems must be driven by
other forms of turbulence. The most favorable bootstrap disk environments appear to be hot, low
density disks with short rotation periods. In such systems collisional damping is relatively weak,
and the instability may be driven by anisotropic pressure and viscosity at global scales.
On the other hand, because the important parameter is the ratio of
cyclotron frequency for the more massive charge carriers to the
rotation frequency, proto-stellar accretion disks in which dust grains
carry most of the charge may exhibit some of the effects we have pointed out.

\acknowledgements
The authors are happy to acknowledge the hospitality of
the KITP at UC Santa Barbara, where this work began. We  thank S. Balbus, V.V. Mirnov, D. Neufeld,
P. Sharma, and J. Stone for useful discussions. This work was supported in part by the National
Science Foundation under Grants PHY99-07949, PHY02-15581, AST05-07367, AST02-05806, and AST-0313031.
%%%%%%%%%%%%%%%%%%%%%%%%%%%%%%%%%%%%%%%%%%%%%%%%%%
%
%\references
%
%%%%%%%%%%%%%%%%%%%%%%%%%%%%%%%%%%%%%%%%%%%%%%%%%%

%%%%%%%%%%%%%%%%%%%%%%%%%%%%%%%%%%%%%%%%%%%%%%%%%%
%%%%%%%%%%%%%%%%%%%%%%%%%%%%%%%%%%%%%%%%%%%%%%%%%%
\end{document}